\documentclass[preprint2]{aastex}
\usepackage{graphicx}
\usepackage{amsmath}
\usepackage{amssymb}
\usepackage{cases}
\usepackage{paralist}
\usepackage{bbold}

\DeclareMathAlphabet{\mathpzc}{OT1}{pzc}{m}{it}

\shorttitle{ACICA algorithm}
\shortauthors{Waldmann}

\begin{document}

\title{On signals faint and sparse: the \texttt{ACICA} algorithm for blind de-trending of Exoplanetary transits with low signal-to-noise }

\author{I. P. Waldmann}
\affil{University College London, Gower Street, WC1E 6BT, UK}
\email{ingo@star.ucl.ac.uk}

\begin{abstract}

Independent Component Analysis (ICA) has recently been shown to be a promising new path in data analysis and de-trending of exoplanetary time series signals. Such approaches do not require or assume any prior or auxiliary knowledge on the data or instrument in order to de-convolve the astrophysical light curve signal from instrument or stellar systematic noise. These methods are often known as `blind source separation' (BSS) algorithms. Unfortunately all BSS methods suffer from a amplitude and sign ambiguity of their de-convolved components which severely limits these methods in low signal-to-noise (S/N) observations where their scalings cannot be determined otherwise. Here we present a novel approach to calibrate ICA using sparse wavelet calibrators. The Amplitude Calibrated Independent Component Analysis (\texttt{ACICA}) allows for the direct retrieval of the independent components' scalings and the robust de-trending of low S/N data. Such an approach gives us an unique and unprecedented insight in the underlying morphology of a data set, making this method a powerful tool for exoplanetary data de-trending and signal diagnostics.

\end{abstract}


\section{Introduction}

As we explore smaller and smaller extrasolar planet around ever fainter stars, it is unsurprising that the need for ever more accurate data-calibration and de-trending techniques is a growing one. In the recent past, there has been a notable emergence of so called `non-parametric' data de-trending algorithms in the fields of transiting extrasolar planet and time-resolved exoplanetary spectroscopy \citep{carter09,thatte10,gibson12,waldmann12b, waldmann13}. The use of such `non-parametric' algorithms is a reactionary response to the difficulties of calibrating and de-trending time series observations when the instrument response function is not known at the precision of the science signal to be extracted. Previous, more conventional `parametric' de-trending approaches rely on auxiliary information of the instrument (e.g. temperature sensor readings, telescope tilt, drifts in x and y positions of the science signal across the detector, etc.). Such methods have the disadvantage of being heavily reliant on the signal-to-noise (S/N) of the auxiliary information used to de-trend the science data, as well as suffering from a degree of arbitrariness in their definition of the instrument response function. 

In \citet{waldmann12b} and \citet{waldmann13}, we have demonstrated independent component analysis (ICA) as novel de-correlation strategy for exoplanetary time series. ICA belongs to the class of blind-source separation (BSS) algorithms, which attempt to de-correlate an observed mixture of signals into its individual source components without prior knowledge of the original signals nor the way they were mixed together. Such an approach requires the least amount of information on a given system and hence ensures a high degree of objectivity in the de-trending of data.

\subsection{Limitations of conventional ICA}
\label{sec:limitations}
Whilst it has been shown that ICA is well suited to the de-correlation of non-Gaussian signals in simultaneously observed exoplanetary time series, it has two mayor limitations that will be addressed in this paper. These are: 

 {\it 1) Susceptibility to Gaussian noise}: to de-correlate non-Gaussian signals, ICA is inherently limited to a low degree of Gaussian white noise in the observed time series observations. This so far posed a significant limitation on the types of data that can be de-correlated. \citet{waldmann13} showed that medium to high-SNR space based observations are somewhat permissible but noisier ground-based observations of exoplanetary time series are often out of reach for conventional ICA algorithms. 
 
 {\it 2) Amplitude and Sign ambiguity}: Like all blind source separation (BSS) algorithms, ICA can de-correlate signals up to an amplitude and sign ambiguity. As explained in section~\ref{sec:bssintro}, the algorithm attempts to simultaneously estimate the source signals, ${\bf s}$, as well as their respective mixings (the mixing matrix), ${\bf A}$ that represent our observations ${\bf x} = {\bf A^{-1} s}$. Given both ${\bf s}$ and ${\bf A}$ are unknown, a scalar multiplication of either can be canceled by an equal division of the other. Hence no BSS algorithms attempt the retrieval of scalar amplitudes of the source signals ${\bf s}$. \citet{waldmann13} resolved this by iteratively fitting components of ${\bf s}$  onto observed out-of-transit data to retrieve the lost scaling factors. Whilst this is a valid approach, it again limits us to high-SNR observations as too much scatter in the observed time-series inhibits a good convergence of such a scaling factor regression. 

In this paper we will address both these limitations by defining ICA in orthogonal wavelet space. In the wavelet domain, as explained in later sections, we can threshold Gaussian wavelet coefficients and increase the signal's sparsity, making the ICA algorithm more robust in difficult S/N conditions. We can furthermore inject a sparse wavelet coefficient calibration signal that allows us to directly calibrate the amplitudes of the mixing matrix {\bf A} without the need of any post-ICA scaling factor regression. 

A quick introduction to BSS and Wavelets are given in sections~\ref{sec:bssintro} \& \ref{sec:waveletintro}, a description of the Wavelet-ICA and noise thresholding in section~\ref{sec:waveletica}. Section~\ref{sec:calibration} describes the amplitude calibration algorithm which is demonstrated in sections~\ref{sec:simulations}~\&~\ref{sec:data} using simulations and {\it Spitzer}/IRS data respectively.

\subsection{Blind Source Separation}
\label{sec:bssintro}
Besides ICA, other BSS algorithms include principal component analysis (PCA), factor analysis (FA), projection pursuit (PP), non-negative matrix factorisation (NMF), stationary subspace analysis (SSA), morphological component analysis (MCA) amongst others. For an extensive review of these algorithms we refer the interested reader to \citet{icabook2} as well as relevant ICA literature \citep{oja92,hyvarinen99,hyvarinen00,icabook,icabook3, koldovsky06, koldovsky05,yeredor00,tichavsky08}. Whereas the underlying statistical assumptions differ significantly, all these algorithms take $N$ simultaneously observed signals $x_{k}(t)$, where $k$ is the observed signal index, and de-correlate these into the source signals $s_{l}(t)$, where $l$ is the source signal index. They all follow the functional form

\begin{align}
\label{intro1}
x_{k}(t) &=  a_{k,1} s_{1}(t) +a_{k,2} s_{2}(t) +  a_{k,3} s_{3}(t) +\\\nonumber
& + a_{k,4} s_{4}(t)  + ... + a_{k,l} s_{N}(t).
\end{align}

\noindent where  $a_{k,l}$ are scaling factors. Assuming that the exoplanetary observation consists of a mixture of astrophysical signal, $s_{a}(t)$, instrument or stellar systematic noise, $s_{sn}(t)$, and the white noise signal, $s_{wn}(t)$, from a Gaussian process $wn(t)$, we can express equation~\ref{intro1} as sum of vectors (the time-dependance has been dropped for clarity):
\begin{equation}
{x}_{k} = a_{k,1}{s}_{a} +  a_{k,2}{s}_{wn} + \sum_{l=3}^{N_{sn}} a_{k,l} {s}_{sn} 
\label{intro2}
\end{equation}

\noindent where $N_{sn}$ is the number of systematic noise sources. Finally this can also be expressed as column vectors ${\bf x} = [x_{1}(t), x_{2}(t), \dots, x_{k}(t)]^{T}$, \\ ${\bf s} = [s_{1}(t), s_{2}(t), \dots, s_{l}(t)]^{T}$ and the mixing matrix ${\bf A}$,

\begin{equation}
\bf{x}=\bf{A}\bf{s}.
\label{intro3}
\end{equation}

\noindent We can furthermore define the `de-mixing matrix' ${\bf W}$ as the inverse of the `mixing matrix'

\begin{equation}
\label{equ:demix}
{\bf W} = {\bf A^{-1}}.
\end{equation}

\noindent For a perfect de-correlation of the observed data ${\bf x}$ into its source components, the original mixing matrix is the perfect inverse of the estimated de-mixing matrix ${\bf W}$ and ${\bf WA = I}$, where ${\bf I}$ is the identity matrix.

ICA attempts to estimate both ${\bf s}$ and the de-mixing matrix ${\bf W}$ by assuming that all components of ${\bf s}$ are statistically independent of one another. This is achieved by iteratively maximising the non-Gaussianity of each signal component by estimating their respective Shannon entropies \citep{shannon48,hyvarinen00}. For more information we refer the reader to \citet{waldmann12b} and the standard literature.

\subsection{Introduction to Wavelets}
\label{sec:waveletintro}

Readers familiar with wavelet decompositions may jump to section~\ref{sec:waveletica}.

Similar to a Fourier Transform (FT), the Wavelet Transform (WT) decomposes a given time series signal into its frequency components. Where the FT uses sine and cosine functions that extend over the full range of the data, the WT uses highly localised impulses. These impulses or `wavelets' scan through the time series and much like a tuning fork to an instrument, `resonate' with localised features in the time series that are akin to the wavelet's shape and scaling. The individual wavelet basis functions are derived from a single mother wavelet $\psi (t)$ through translation and dilation of the mother wavelet \citep{percival00}. Different wavelets exist with different analytical properties, here we use the orthogonal basis wavelets of the Daubechies (db) family \citep{daubechies88}. The wavelet analogue to the Fourier transform of the times series $x(t)$ is given by:

\begin{equation}
c_{\tau,\varphi} = \int_{\mathbb{R}} x(t) \psi_{\tau,\varphi} (t) \text{d}t
\label{wavelet1}
\end{equation}

\noindent where $ \psi_{\tau,\varphi} (t)$ is called the `mother wavelet' for a given scaling $\varphi$ and translation $\tau$ and $c$ is the wavelet coefficient with respect to $\tau$ and $\varphi$. We define the mother wavelet for the continuous wavelet transform (CWT) as

\begin{equation}
\psi_{\tau,\varphi} = \frac{1}{\sqrt{2}} \psi \left( \frac{t-\tau}{\varphi} \right)
\label{motherscale}
\end{equation}

 The wavelet base is orthogonal and we can hence reconstruct the data by taking the sum of the product of all coefficients for a given scale and translation, $c_{\tau,\varphi}$, with the respectively scaled and translated mother wavelet 

\begin{equation}
x(t) = \sum_{\varphi \in \mathbb{Z}} \sum_{\tau \in \mathbb{Z}} c_{\tau,\varphi} \psi_{\tau,\varphi}(t)
\label{wavelet2}
\end{equation}

For a more in-depth definition of wavelets and their respective properties we refer the interested reader to \citet{daubechies92} and \citet{percival00}.

\subsubsection{Multi-resolution analysis}

The above equations apply to the CWT case. The wavelet coefficients describe the correlation between the wavelet at varying scales (or frequencies). These can be calculated by changing the scale of the wavelet, i.e. the analysis window. We can hence speak of a multi-resolution decomposition, where each scaling of the mother wavelet denotes a given resolution. Here, the analogy to the Fourier Transform would be band-pass filters of varying size. 
Most times it is more sensible to exploit the discrete nature of the data and to define the discrete wavelet transform (DWT). The DWT is significantly easier to implement and faster to compute. 
Similarly to the continuous case, in the DWT we have a `mother' wavelet and a scaling function, also known as `father' wavelet. Here, the `mother' wavelet is denoted by $h(t)$ and the `father' by $g(t)$ \citep{daubechies92, percival00,press07}.  It is important to note that unlike in the CWT case, where the `mother' wavelet itself is scaled to represent different frequencies in the data, this is not the case in the DWT. In the DWT, in analogy with the Fourier Transform, $h(t)$ and $g(t)$ can be thought of as high-pass and low-pass frequency filters respectively. Different scalings are then achieved by progressively `down-sampling' the data.

\begin{figure}[t]
\centering
\includegraphics[width=7cm]{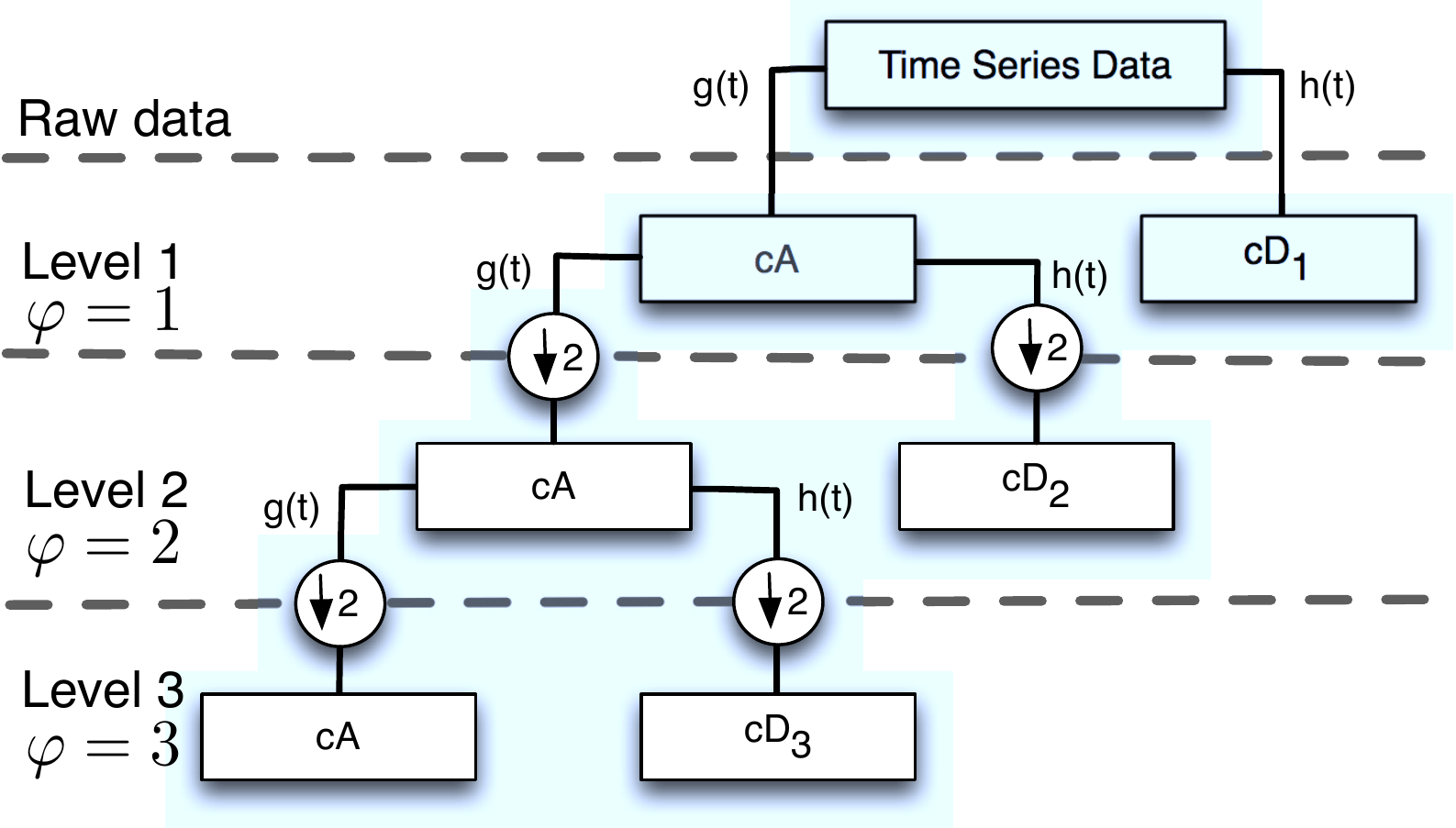}
\caption{Outline of multi-resolution wavelet decomposition down to the 3rd decomposition level. \label{waveletchart}}
\end{figure}

\begin{figure}[t]
\centering
\includegraphics[width=7cm]{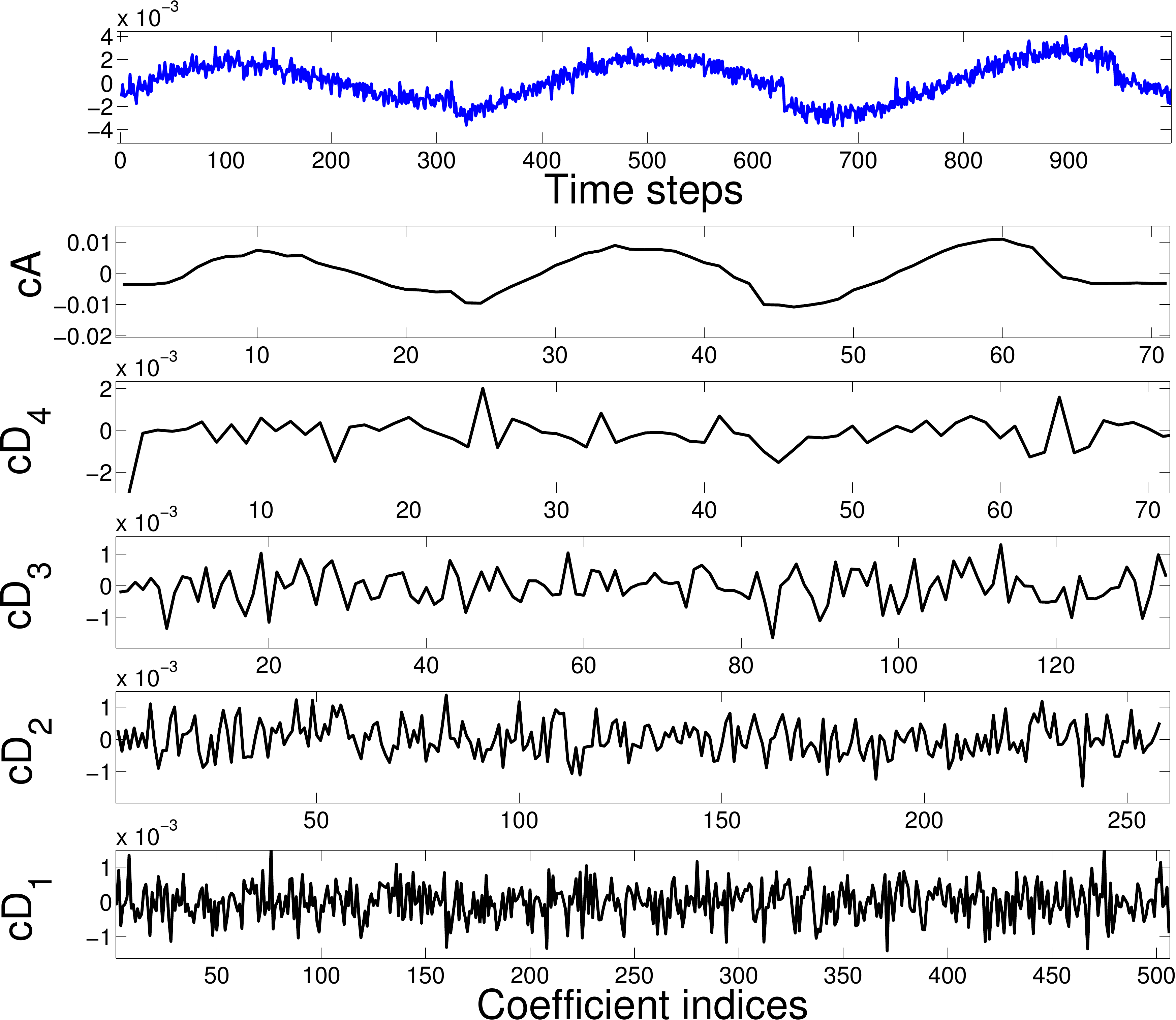}
\caption{Example of a discrete wavelet transform (DWT). TOP: sinusoidal time series with Gaussian noise and saw-tooth functions superimposed. BOTTOM: four level DWT decomposition of a noisy sinusoid using {\it symlet-5} wavelets. It can be seen that the average coefficients, $cA$, represent a `moving average' of the data, whereas the detailed coefficients, $cD_{\varphi}$, represent bands of higher frequencies.\label{dwtexample}}
\end{figure}

The DWT is best understood by following the individual steps of the algorithm that computes the transform:

\begin{enumerate}
\item The observed, discrete time series, $x(t)$, is convolved with the `mother' wavelet $h(t)$:

\begin{equation}
cD_{\varphi}(t)= (x \ast h)(t) = \sum_{\tau=-\infty}^{\infty} x(\tau) \cdot h(t-\tau).
\label{dwt1}
\end{equation}

\noindent where $(x \ast h)$ denotes the convolution of $x$ with $h$ and $cD_{\varphi}$ represent the `mother' wavelet coefficients for a given scale $\varphi$. As mentioned earlier, the `mother' wavelet, $h(t)$ acts as a high-pass filter, sensitive to the high frequencies or details of the time series. We hence refer to the coefficients of $h(t)$ as {\it detail coefficients}. 

\item The next step is to convolve the same time series, $x(t)$, with the scaling function or `father' wavelet:

\begin{equation}
cA_{\varphi}(t) = (x \ast g)(t) = \sum_{\tau=-\infty}^{\infty} x(\tau) \cdot g(t-\tau).
\label{dwt1-2}
\end{equation}

\noindent As opposed to the `mother' wavelet, the `father' wavelet acts as a low-pass filter of the time series and its coefficients can be viewed as a moving average of the underlying trend of $x(t)$. We hence refer to its coefficients as {\it average coefficients} and denote them with $cA_{\varphi}$. Furthermore, the low-pass filter $g(t)$ is related to the high-pass filter by 

\begin{equation}
g(L-1-t)~=~(-1)^{t}~\cdot~h(t) 
\end{equation}

\noindent where $L$ is the filter length and corresponds to the number of points in the time series $x(t)$.

\item We now have two sets of time series, a low-pass filtered, moving-average time series, $cA_{\varphi}$, and a high-pass filtered time series, $cD_{\varphi}$. We record the $cD_{\varphi}$ coefficients and proceed with our analysis of the average coefficients, $cA_{\varphi}$. Since half of the frequencies in $cA_{\varphi}$ (namely the high-pass ones) have been removed by equation~\ref{dwt1-2}, the Nyquist theorem tells us that we are oversampled by a factor of two. We can hence remove every second coefficient in $cA_{\varphi}$ without losing information. This operation is termed `down-sampling' and abbreviated by $\downarrow 2$, leaving us with $N/2$ coefficients to describe $cA_{\varphi}$. Similar applies to the detailed coefficients $cD_{\varphi}$. The detail and average coefficients are hence given by:

\begin{equation}
cA_{\varphi}(\tau) =  \sum_{t} x(t) \cdot g(2\tau-t)
\label{dwt2}
\end{equation}

\begin{equation}
cD_{\varphi}(\tau) =  \sum_{t} x(t) \cdot h(2\tau-t)
\label{dwt3}
\end{equation}

\item The Nyquist down-sampling introduces the concept of scaling or multiple resolutions. If we now repeat steps 1-3 on the down-sampled average coefficients, $cA_{\varphi}$, we obtain a new set of coefficients ($cA_{\varphi+1}$ and $cD_{\varphi+1}$) on a scale that is double the size of the previous decomposition. This is illustrated in figure~\ref{waveletchart} as flow chart and a data example is given in figure~\ref{dwtexample}.

\end{enumerate} 

\noindent For a given scale, $\varphi$, the data can now be reconstructed by reversing the above process:

\begin{align}
\label{dwtrecon}
x_{\varphi}(t) &= (cA_{\varphi=\Phi}(\tau) \cdot g(-t+2\tau)) \\\nonumber
&+ \sum_{\varphi}\sum_{\tau=-\infty}^{\infty}  (cD_{\varphi} \cdot h(-t + 2\tau))
\end{align}

\noindent where $\Phi$ are the total number of decompositions. 

For an algorithmic implementation using quadrature mirror filters (QMFs) see appendix~\ref{sec:QMF} and \citet{press07}.


\section{Wavelet ICA}
\label{sec:waveletica}

We now perform the DWT on each observed time series, $x_{k}$, and obtain a series of average coefficients, $cA_{k}$, and detail coefficients for a given scale, $cD_{k,\varphi}$. For our ICA decomposition, we use these series of coefficients instead of the time domain time-series, $x_{k}$, and define our observed data as 

\begin{equation}
\hat{x}_{k} = \sum_{\tau}cA_{k}(\tau) + \sum_{\varphi}^{\Phi}\sum_{\tau} cD_{k,\varphi}(\tau)
\label{equ:waveleticaobs}
\end{equation}

\noindent and similarly we can express our source signals, $s_{l}$, as the wavelet equivalent

\begin{equation}
\hat{s}_{l} = \sum_{\tau}cA_{l}(\tau) + \sum_{\varphi}^{\Phi}\sum_{\tau} cD_{l,\varphi}(\tau).
\end{equation}

\noindent From here onwards we will use $\hat{x}$ to denote the wavelet domain presentation of the time-domain signal $x$. We hence restate equation~\ref{intro3} as 

\begin{equation}
\hat{\bf x} = \hat{\bf A} \hat{\bf s}.
\label{equ:waveica}
\end{equation} 

There are several important considerations to note here: 

As we are dealing with a multi-resolution analysis, it becomes possible to perform the blind source separation on individual scales (or bandpasses) only or to actively exclude some frequency ranges from the analysis. Such an approach may be advantageous if one has prior knowledge of the signals' frequency bandwidths and wishes to restrict the impact of high-frequency noise or other signals on the BSS of a given signal. An example of this is given by \citet{lin05}. In this paper, we do not take this approach for reasons described in section~\ref{sec:calibration}.

Transforming the observed data into wavelet basis sets increases the redundancy of the data. This is advantageous in the case of overcomplete ICA, where more source signals are present in the data than time series observed. This overcompleteness leads to an improper separation of the independent source signals in the data. Increasing the data's redundancy has been shown to alleviate this problem and makes it possible to efficiently use ICA in very restricted data ranges \citep{inuso07,mammone12}.

Furthermore as pointed out in section~\ref{sec:limitations}, ICA is inefficient in the presence of high frequency scatter. This is alleviated in the wavelet space as by the virtue of the multi-resolution analysis, most high-frequency scatter is contained in the $cD_{1}$ coefficients and a better degree of separation can be achieved for lower frequency systematics. We will demonstrate this property in section~\ref{sec:data}. It is also possible in wavelet space to selectively suppress Gaussian noise via soft or hard thresholding \citep[e.g. ][]{stein81,donoho95} and hence increase the robustness of the BSS algorithm in low S/N conditions.

\subsection{Amplitude Calibrated ICA (\texttt{ACICA})}
\label{sec:calibration}

Arguably the central problem with an exoplanetary data de-trending algorithm based on ICA is the scale and sign ambiguity of the de-correlated, independent components. 

Whilst we do not know the scaling of individual source components in ${\bf s}$, we know by definition of the ICA pre-processing step (whitening) that each source signal is normalised to unit variance, $E[s_{l}^{2}] = 1$, and that ${\bf x} = {\bf A s}$. Hence the elements of ${\bf A}$ do not need to preserve the absolute but the relative scalings of ${\bf s}$. 
 If we know the absolute scaling of one source signal, we can hence solve the scaling degeneracy of all signals. This is usually possible for simulations where we control the inputs but not for real life examples where the source signals and their scalings are {\it a priori} unknown. One way to solve this predicament is via the introduction of a calibration signal (CS). Such a CS must have the following properties: 
 
1) {\it Minimal to no impact on the data}: The introduction of a CS should not distort any underlying signals or their amplitudes.

2) {\it Temporal localisation}:  The CS should not be located in in-transit (INT) regions and not take up too much out-of-transit (OOT) data.

3) {\it Stability to noise}: The CS should not be affected by the noise of the data (be it Gaussian or otherwise).
 
 4) {\it Non-Gaussianity}: The signal should be non-Gaussian enough for the ICA to recognise it but not more non-Gaussian than the science signal itself. A too prominent CS could bias the ICA towards the retrieval of the CS and could impair the retrieval of other non-Gaussian signals.

 In the time domain, it is difficult to implement property 4 without violating property 1. In order to implement a distinct enough non-Gaussian signal in the time domain, one needs to significantly alter large sections of the data. Also the treatment of noise (property 3) is problematic. In the frequency domain, property 2 is violated as the Fourier transform does not contain temporal information and the CS would superimpose the science signal. However, all four criteria can be met in the wavelet domain. 

\subsubsection{Injecting the Calibration Signal}

In the wavelet domain, we can introduce a much sparser CS than in the time domain and have control over its temporal location (unlike in the frequency domain) via the lag term $\tau$. Given that $\hat{\bf x}$ is more redundant than ${\bf x}$, we can minimise the impact on the observed data or avoid any alterations to the data altogether by selecting wavelet coefficients of $\hat{\bf x}$ with zero or near-zero amplitudes to be used for our CS. We can now re-define equation~\ref{intro2} as

\begin{equation}
\label{equ:ICArefsignal}
\hat{x}_{k}^{c} = \hat{x}_{k} + \hat{b}_{k} \hat{s}_{c} =   \hat{a}_{k,1}\hat{s}_{c} + \sum^{N}_{l=2} \hat{a}_{k,l} \hat{s}_{l} 
\end{equation}

\noindent where $\hat{x}_{k}^{c}$ denotes the observed data with CS, $\hat{s}_{c}$ is the CS with the same dimensions as $s_{l}$ and $\hat{b}_{k}$ is a random but known scaling constant. We define $\hat{s}_{c}$ as 


\begin{equation}
\label{equ:calibsignal}
\hat{s}_{c} = \left\{ 
  \begin{array}{l l}
  cD_{\varphi}(\tau) = 1  &  \text{if~} \tau = \tau_{\varphi}^{c}\\
    \sum_{\tau}cA(\tau)+ \sum_{\varphi}^{\Phi}\sum_{\tau} cD_{\varphi}(\tau) = 0  &  \text{otherwise}\\
  \end{array} \right.
\end{equation}

\noindent where $\tau_{\varphi}^{c}$ are pre-selected lags for a given scale that correspond to a section of the OOT data. Note that: 

1) No average coefficients, $cA$, are used in $\hat{s}_{c}$ since these are not redundant enough. 

2) As equation~\ref{equ:ICArefsignal} suggests, only $N$ number of independent components can be extracted. Adding the CS in the the observed data $\hat{\bf x}$ may render the ICA overcomplete as one source signal will not be retrieved in favour of the CS. For large data sets ($ N_{observations} > N_{signals}$) this is generally not a problem.

3) We choose $0 <\hat{b}_{k} < \frac{1}{2} \max |\hat{x}_{k}| $ to avoid the CS to be the most dominant feature in the data. 

4) In equation~\ref{equ:calibsignal} we have chosen to define $\hat{s}_{c}$ to contain one non-zero coefficient per scale $\varphi$ and lags corresponding to the same area of OOT data in the time-domain. This allows for efficient use of OOT data and a high sparsity in the wavelet domain but is entirely arbitrary otherwise. 

\subsubsection{Retrieving the scaling information}
\label{sec:scaling}

Having injected the calibration signal into our observations, we can now perform the ICA deconvolution (section~\ref{sec:bssintro}) on the data with CS in the wavelet domain, $\hat{\bf x}^{c}$. We identify the CS in the retrieved source signals and denote their respective elements of the mixing matrix, $\hat{\bf A}$ as $\hat{a}_{k}^{c}$. By measuring the average amplitude of the wavelet coefficients comprising the retrieved CS, $\langle \hat{s}_{c} \rangle$, and knowing the original CS amplitude, $\hat{b}_{k}$,

\begin{equation}
\langle \hat{s}_{c} \rangle~ = \frac{\sum^{\Phi}_{\varphi}\sum_{\tau}cD_{\varphi}(\tau)}{\Phi}
\end{equation}

\noindent we can retrieve the scaling of the CS as well as those of the other signals contained in $\hat{\bf s}$. We denote this calibrated mixing matrix as $\hat{\bf O}$ with its elements given by 

\begin{equation}
\hat{o}_{k,l} = \frac{\hat{a}_{k,l}}{\hat{a}_{k}^{c}} \times \frac{\hat{b}_{k}}{\langle \hat{s}_{c} \rangle}.
\end{equation}

\subsubsection{Calibration error}

Using the above calibration approach we consider the total error on the independent components (ICs) to be a combination of source separation error (SSE) of the individual IC and the SSE of the calibration signal components, $s_{c}$. The SSE for the $l^{th}$ source signal can be estimated by

\begin{equation}
\sigma_{l} = \frac{{\text E}[\sum^{N}_{l=1,l \neq k} \textbf{G}^{2}_{kl}]} {{\text E}[\textbf{G}^{2}_{kk}]}, ~ k ,l = 1,2,...,N.
\label{isrk}
\end{equation}

\noindent where ${\bf G}$ is called the `gain matrix' and defined as ${\bf G = WA}$. For perfectly separated sources, we hence have ${\bf G = WA= I}$ where ${\bf I}$ is the identity matrix. The ICA algorithms employed here \citep[EFICA and WASOBI;][]{koldovsky06,yeredor00, tichavsky06b} can be shown to be asymptotically efficient and converge to the correct solution  in the limit of $N_{iter} \rightarrow \infty$ iterations. However in real world scenarios this is not the case and we find that the estimated mixing (or de-mixing) matrix is only approximately equal to the true underlying mixing matrix ${\bf A}$, i.e. ${\bf W \simeq A^{-1}}$. To estimate the SSE, we can hence inspect the variance of the matrix ${\bf G}$. Whilst it is possible to directly calculate ${\bf G}$ for simulations, the true mixing of the signals is usually unknown in real data applications. \citet{tichavsky06,koldovsky06} and \citet{tichavsky08} have shown that an asymptotic estimate of ${\bf G}$ is nonetheless possible. For a derivation we refer the reader to the cited literature. 

We define the source separation error for the calibrated components in ${\bf s}^{c}$ as the quadrature error of the individual source components' error and the SSE of the calibration signal

\begin{equation}
\sigma^{c}_{l \neq c} = \sqrt{\sigma^{2}_{l \neq c} + \sigma^{2}_{c}}
\end{equation}

\section{Application Examples}
\subsection{Simulations}
\label{sec:simulations}

In this section we illustrate the ICA calibration approach using simulations. 

\begin{figure}
\centering
\includegraphics[width=7cm]{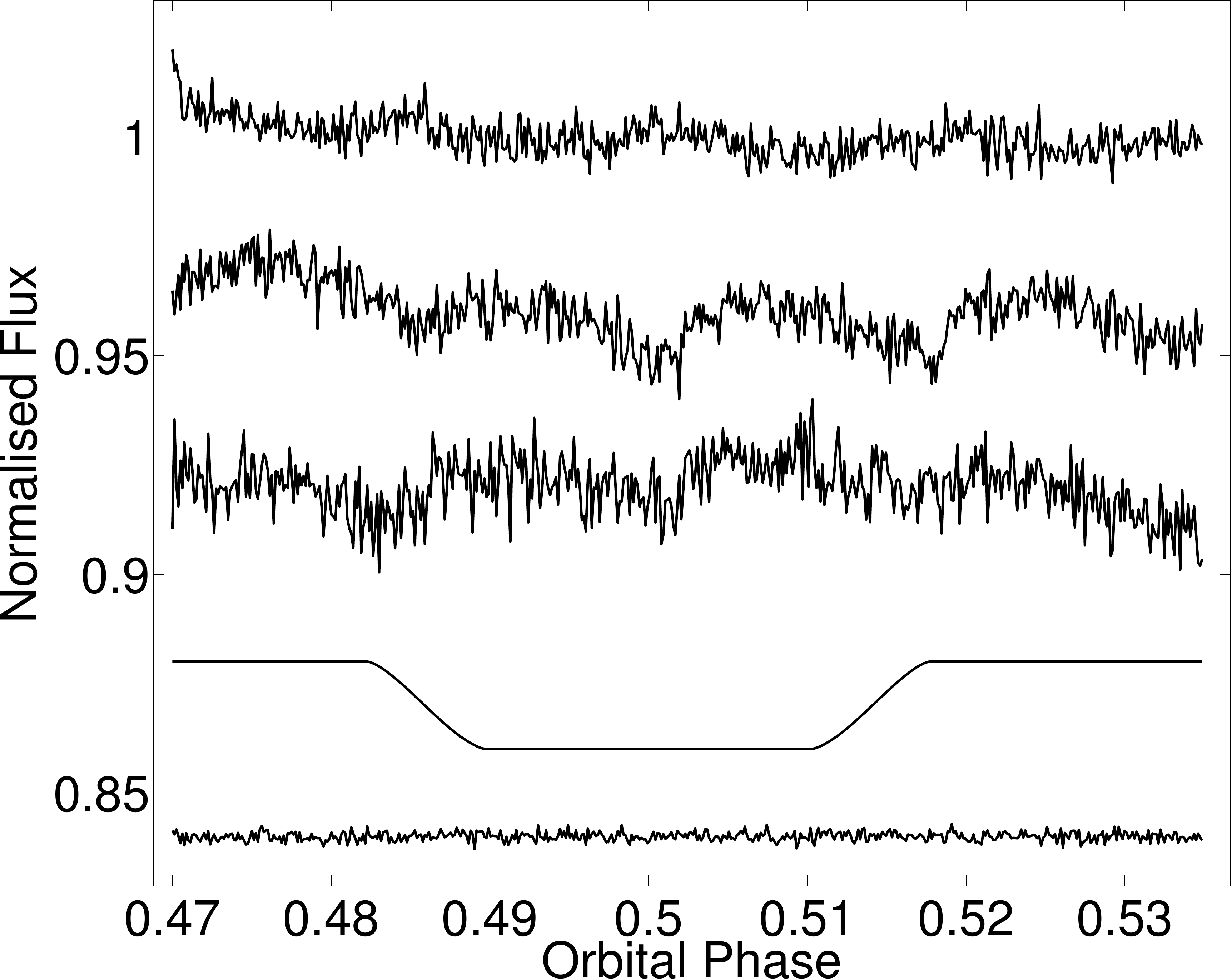}
\caption{Input signals before mixing. From to bottom: 3 systematic noise components of the HST/NICMOS instrument \citep{waldmann13}; a \citet{mandel02} lightcurve of the secondary eclipse of HD189733b; Gaussian noise at with FWHM of 2$\times$10$^{-3}$ normalised flux. \label{diag:simu1}}
\end{figure}

1) For this we generated five input signals: a \citet{mandel02} secondary eclipse curve of HD189733b, a Gaussian process with FWHM of 2$\times$10$^{-3}$ amplitude, and three instrument noise vectors derived from HST/NICMOS observations \citep{waldmann13}. These input signals are shown in figure~\ref{diag:simu1}. 

\begin{figure}
\centering
\includegraphics[width=7cm]{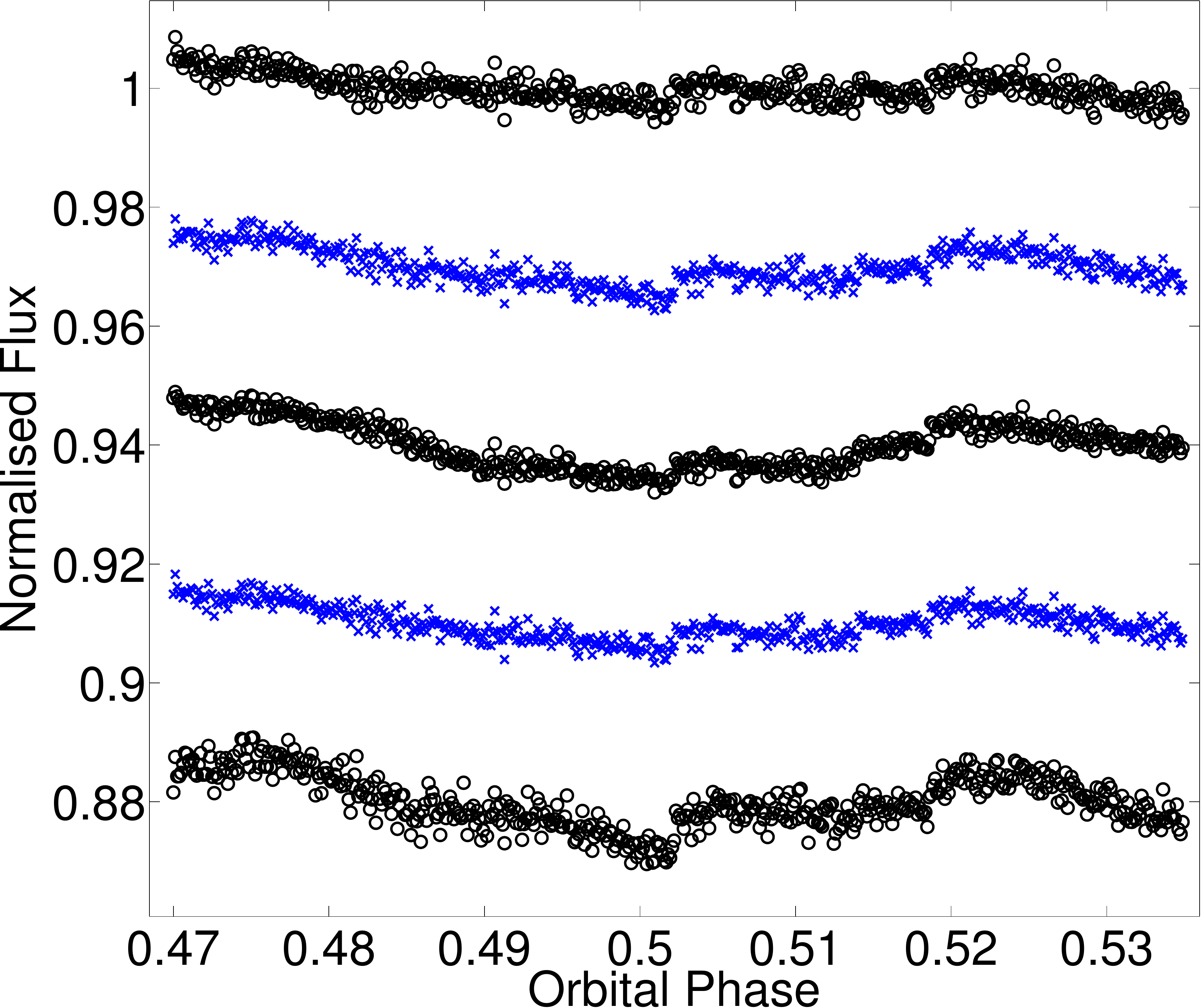}
\caption{Signals from figure~\ref{diag:simu1} mixed together, using a random mixing matrix, to created the observed signals ${\bf x}$. \label{diag:simu2}}
\end{figure}

2) We mix those signals using a randomly generated mixing matrix to obtain our `observed' time series, ${\bf x}$, figure~\ref{diag:simu2}. 

\begin{figure}
\centering
\includegraphics[width=7cm]{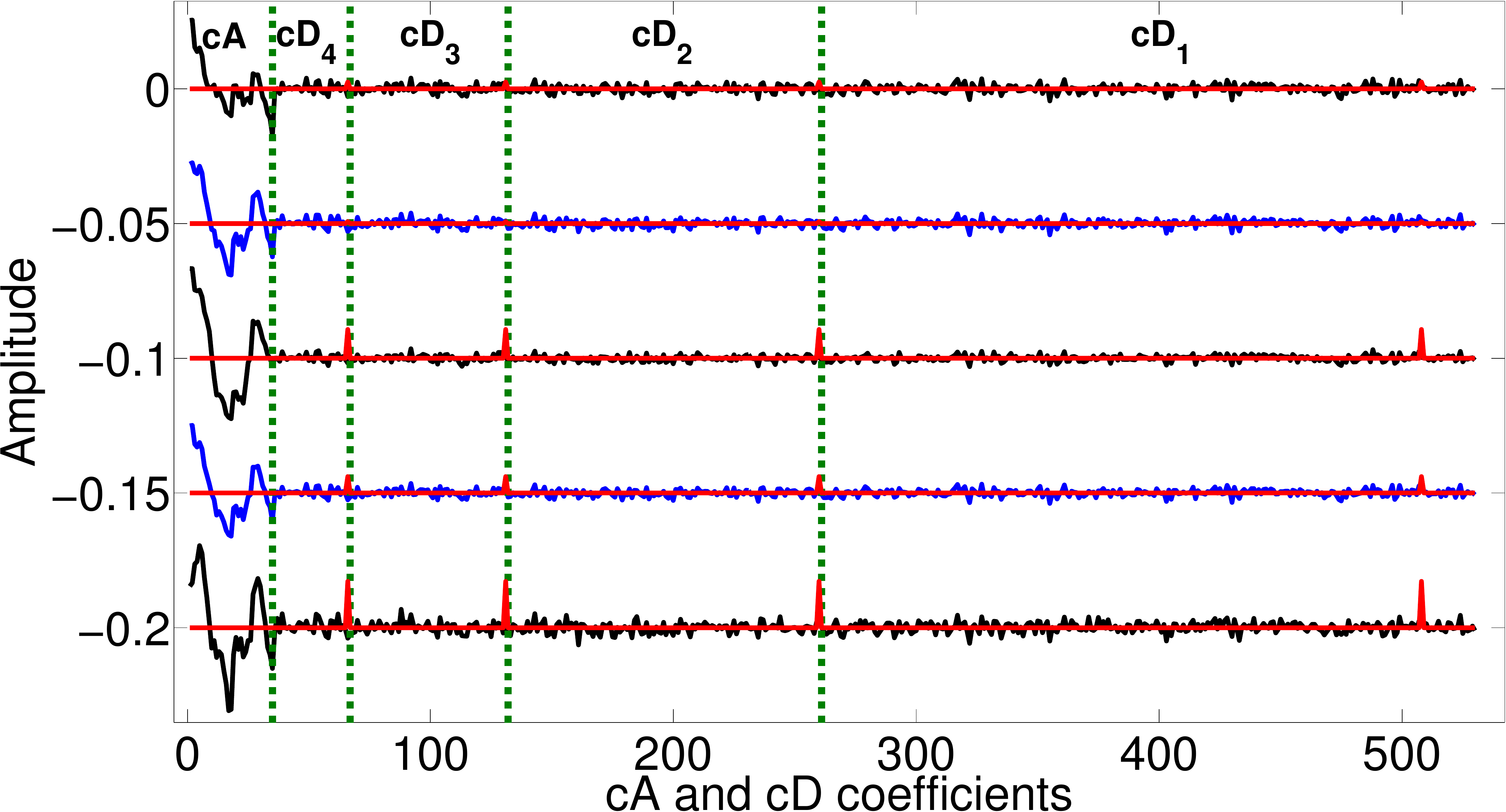}
\caption{BLUE AND BLACK: Wavelet transform of signals in figure~\ref{diag:simu2}, the order corresponds to order in figure~\ref{diag:simu2} (colours for clarity). GREEN dotted: denoting scale boundaries from average coefficients $cA$ to the highest frequency scale $cD_{1}$. RED: Calibration signal for each series $\hat{\bf x}$ over-plotted. Four peaks are visible (one per scale) close to their scale boundaries. Their amplitudes are defined by the random scaling vector $\hat{\bf b}$. \label{diag:simu3}}
\end{figure}

3) Using the \texttt{MATLAB}\footnote{http://www.mathworks.co.uk} wavelet-toolbox and daubechies 4 (db4) wavelets \citep{daubechies88,daubechies92} we calculated the discrete wavelet transform for four scales ($\Phi = 4$) of each time series in ${\bf x}$ to obtain $\hat{\bf x}$ (figure~\ref{diag:simu3}, black and blue curves). In the scope of this simulation we found the choice of wavelet and number of vanishing moments not to matter greatly. However the choice of wavelet and decomposition depth may vary from data set to data set. 

\begin{figure}
\centering
\includegraphics[width=7cm]{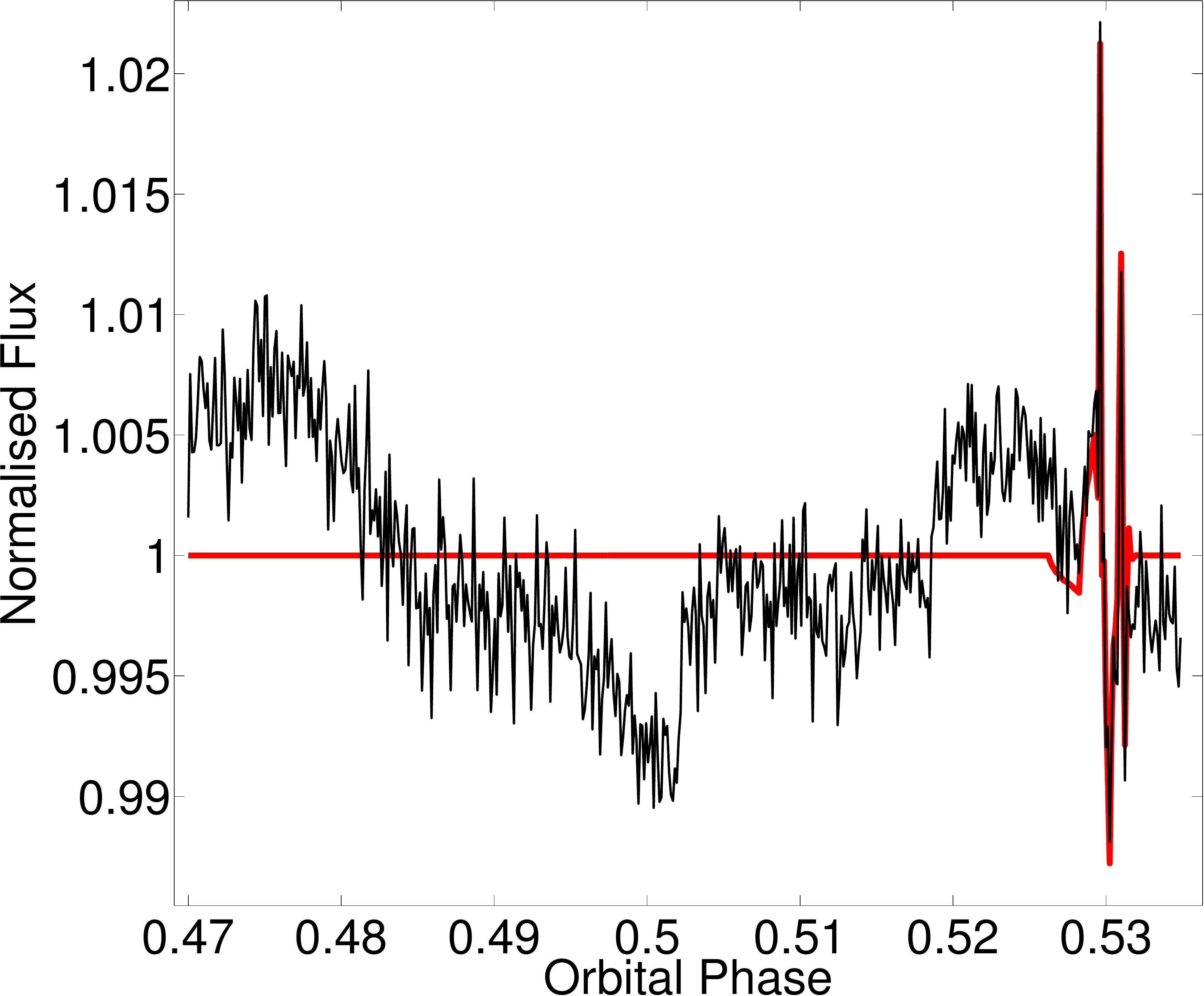}
\caption{BLACK: Time domain representation of third observed component $x_{k=3}^{c}$ after the calibration signal was added (in figure~\ref{diag:simu3}). RED: Calibration signal only. Note the temporal location of the four frequency signal (red spikes in figure~\ref{diag:simu3}) at the far edge of the post-egress out-of-transit data.  \label{diag:simu4}}
\end{figure}

4) We generated the calibration signal, ${s}_{c}$, to consist of one wavelet coefficient per scale $\varphi$ giving $\Phi$ total number of non-zero coefficients. For each series, $\hat{x}_{k}$, the calibration signal was multiplied with the random scaling factor $b_{k}$, i.e. $s_{c,k} = b_{k} s_{c}$.  Figure~\ref{diag:simu3} (red peaks) shows the scaled calibration signal for each $\hat{x}_{k}$. Note that we chose the lags of the non-zero coefficient to correspond to the 95$^{th}$ percentile of each scale. This guarantees the CS to be located in the post-eclipse out-of-transit data. It also overlaps the differently scaled wavelet impulses and hence minimises the impact on the time-domain data representation. See figure~\ref{diag:simu4} for a time-domain representation of $x_{k=3}^{c}$ (third from top in figure~\ref{diag:simu3}).

\begin{figure}
\centering
\includegraphics[width=7cm]{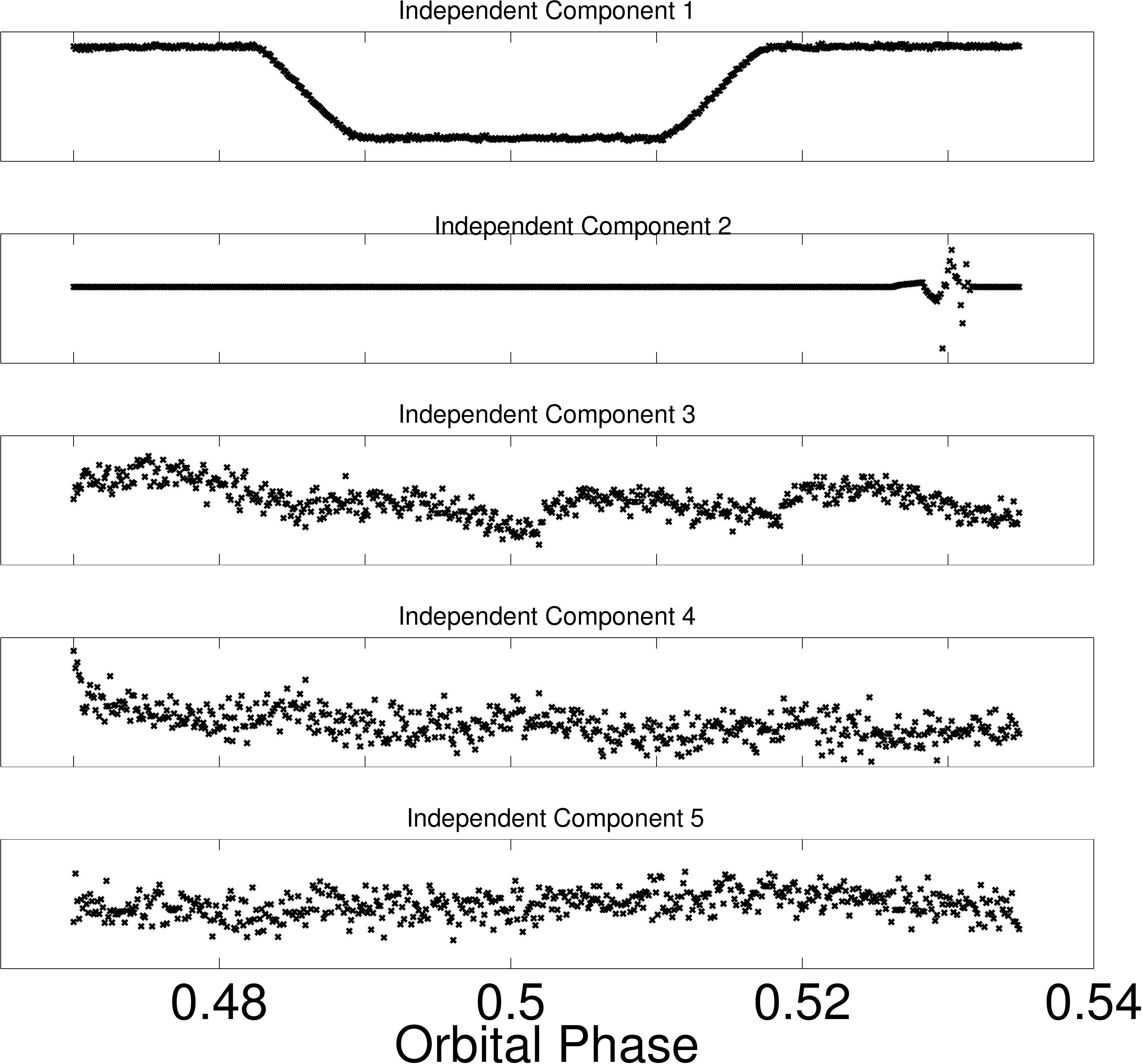}
\caption{Time domain representation of the independent components, ${\bf s}$, retrieved. First components is the secondary eclipse signal, second components the calibration signal. Note that these are not scaled yet. \label{diag:simu5}}
\end{figure}

5) The blind source separation via independent component analysis as described in \citet{waldmann12b} was performed and the time-domain representation of the retrieved ICs are shown in figure~\ref{diag:simu5}. 

\begin{figure}
\centering
\includegraphics[width=7cm]{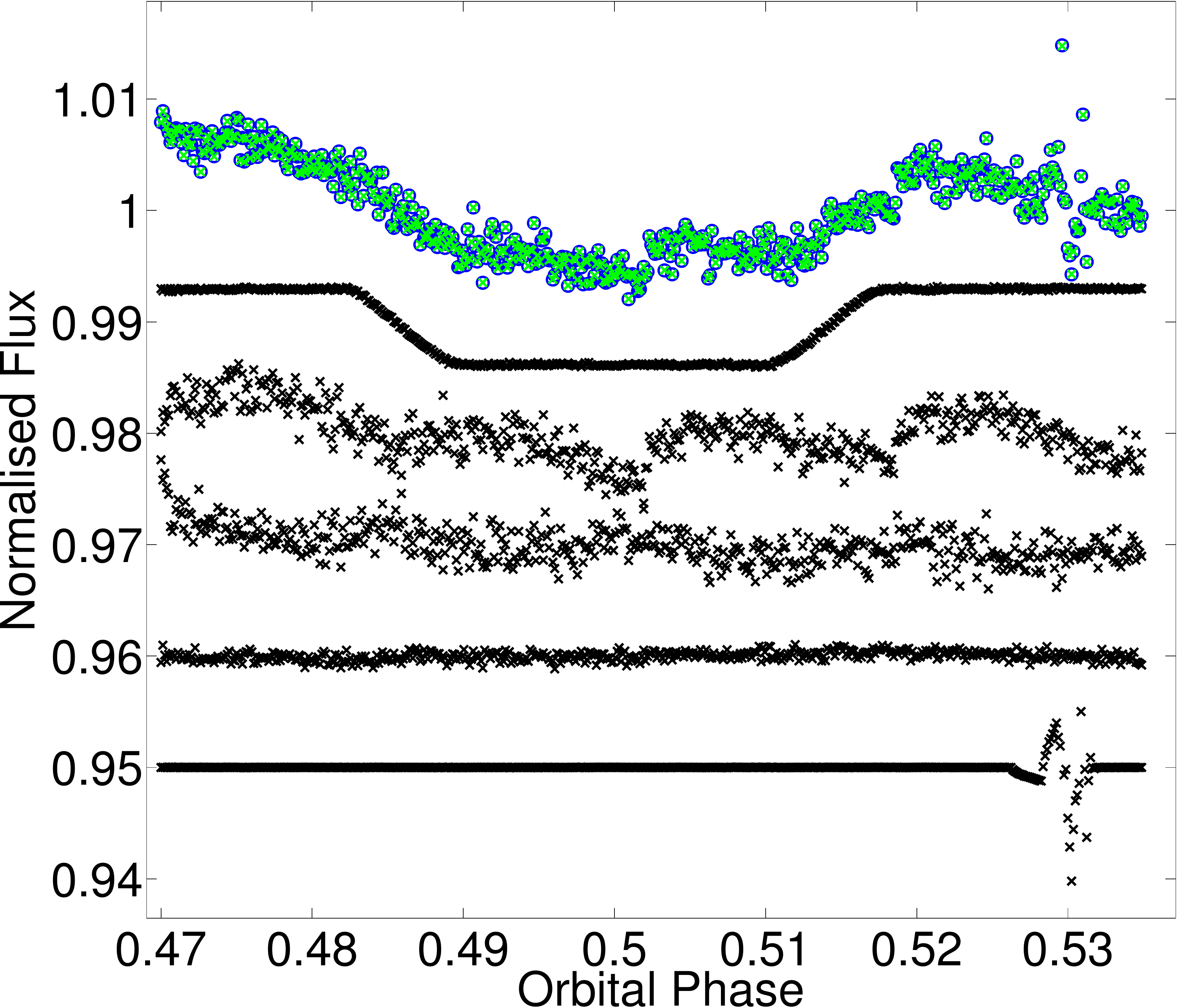}
\caption{TOP: Third observed time series ($x_{k=3}^{c}$, figure~\ref{diag:simu4}, blue circles), over-plotted (green crosses) reconstructed signal using scaled components, described in section~\ref{sec:scaling}. Note the excellent match between observed and reconstructed time series. BOTTOM black: scaled independent components of the above signal.  \label{diag:simu6}}
\end{figure}

6) Following section~\ref{sec:scaling} we calculate the scaled mixing matrix ${\bf O}$ and apply the scalings to individual source signals. Figure~\ref{diag:simu6} shows the observed data $x^{c}_{k=3}$ from figure~\ref{diag:simu4}, blue circles, and the re-constructed data using the scaling matrix ${\bf O}_{k=3,l}$, green crosses. The scaled ICs are shown offset underneath. 

\subsection{Spitzer/IRS: HD189733b secondary eclipse}
\label{sec:data}

\begin{figure}
\centering
\includegraphics[width=7cm]{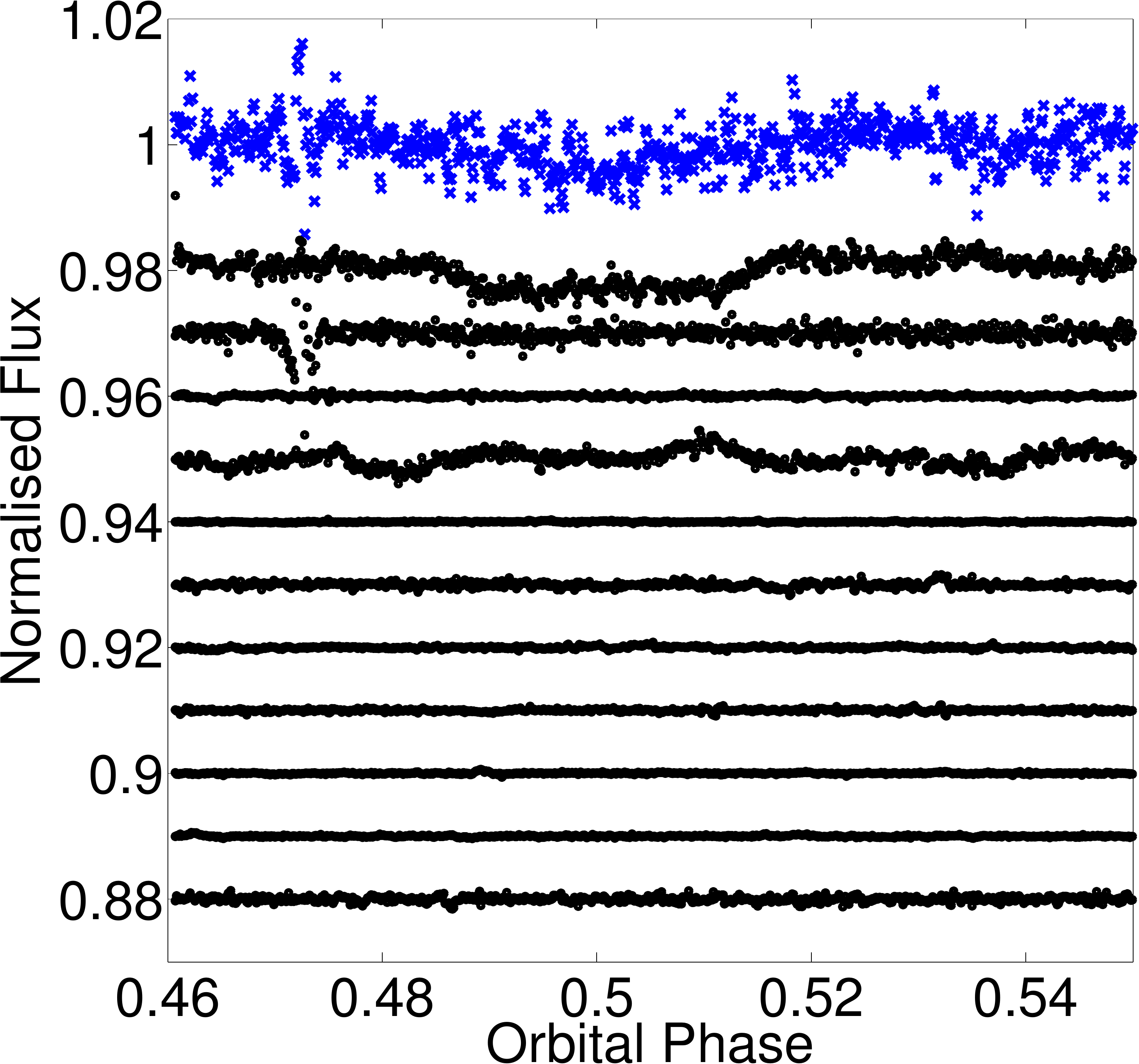}
\caption{Spitzer/IRS observations of HD189733b at 7.6971$\mu$m. BLUE CROSSES: Raw time series. BLACK DOTS: first 11 most non-Gaussian independent components comprising the raw time series. The first and second components are the secondary eclipse curve and the calibration signal respectively. Remaining components are instrument or stellar systematics. \label{diag:spitzer1}}
\end{figure}

\begin{figure}
\centering
\includegraphics[width=7cm]{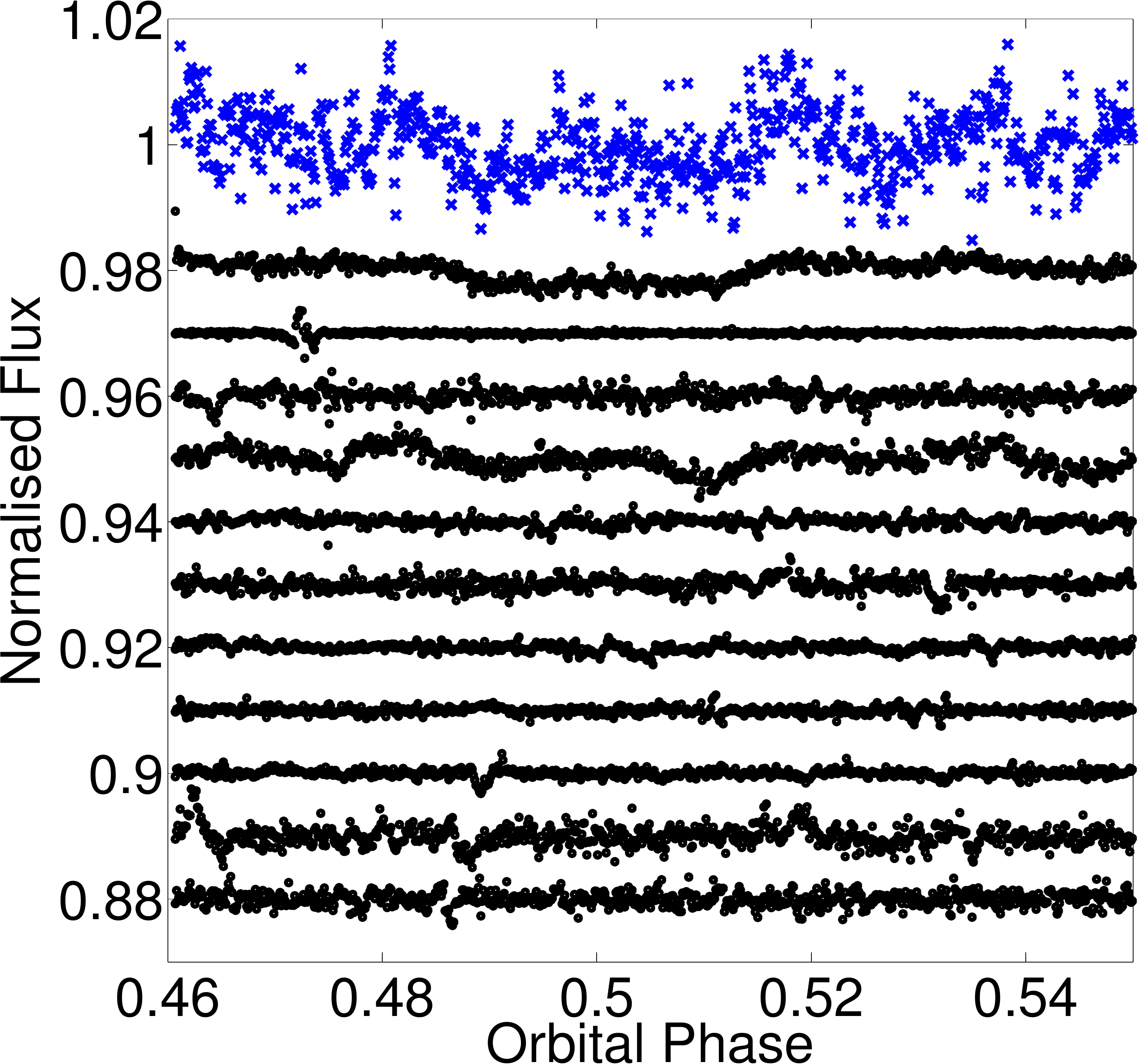}
\caption{Spitzer/IRS observations of HD189733b at 11.6285$\mu$m. BLUE CROSSES: Raw time series. BLACK DOTS: individual independent components in the same order as in figure~\ref{diag:spitzer1}. Note the significantly higher systematic noise amplitudes at the red-end of the spectrum compared to the blue end in figure~\ref{diag:spitzer1}. \label{diag:spitzer2}}
\end{figure}

\begin{figure}
\centering
\includegraphics[width=7cm]{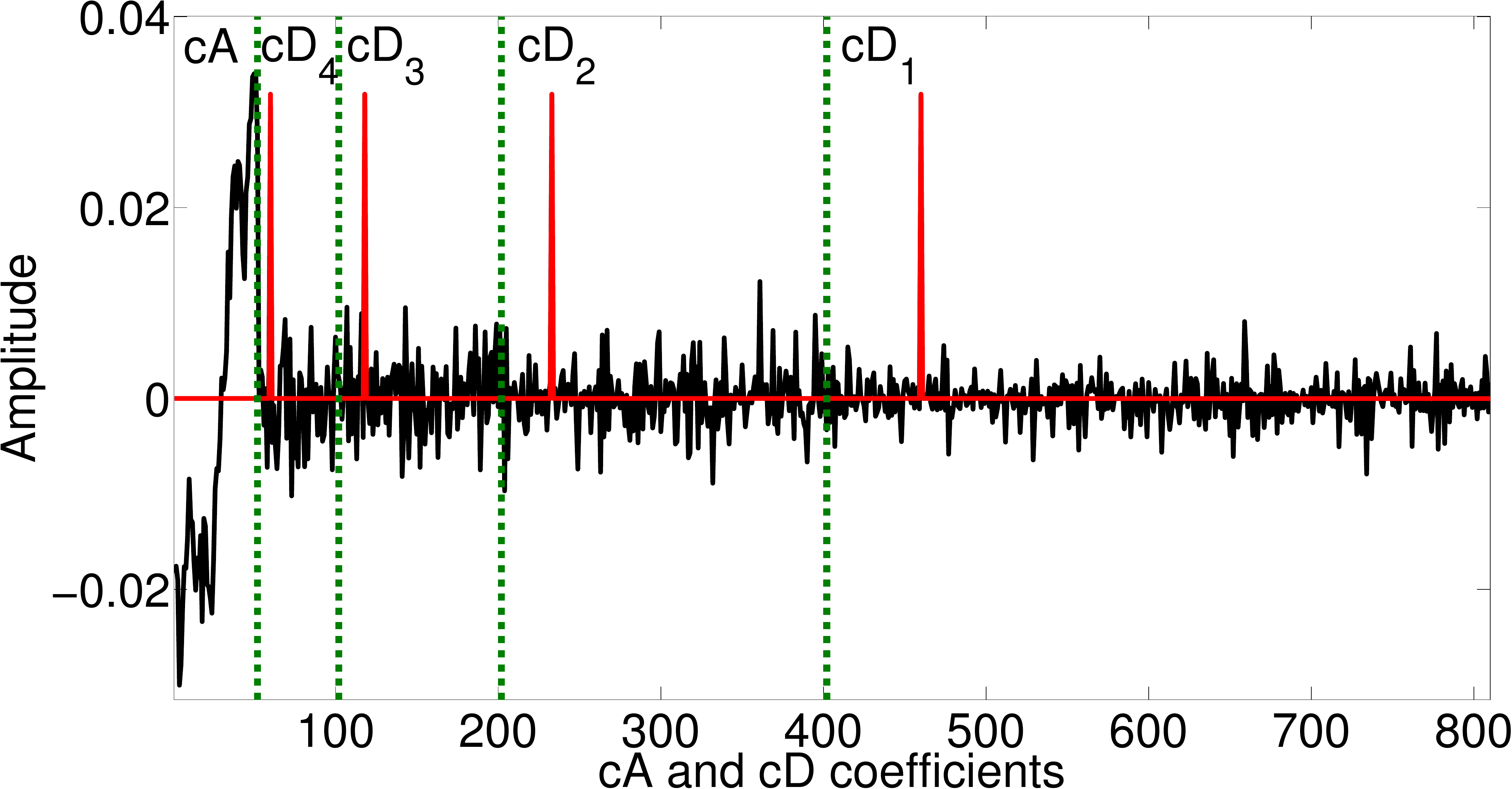}
\caption{BLACK: Wavelet transform presentation of time series in figure~\ref{diag:spitzer1} at 7.6971$\mu$m.  RED: Calibration signal injected in the detailed coefficients. \label{diag:spitzer3}}
\end{figure}

We test the proposed algorithm on Spitzer/IRS \citep{houck04} observations of a secondary eclipse of HD189733b. These observations were obtained in November 2006 (program id: 30473) in low resolution mode ranging from 7.46 to 14.29$\mu$m. The secondary eclipse was followed for 5:48 hours with integration times of 14.7 seconds per ramp. The Spitzer pipeline calibrated data were reduced using the \texttt{Spice}\footnote{version 2.5. http://irsa.ipac.caltech.edu/data\\/SPITZER/docs/dataanalysistools/tools/spice/} spectral reduction software. Examples of the resulting time series are shown in figures~\ref{diag:spitzer1}~\&~\ref{diag:spitzer2} (blue crosses). Similar to section~\ref{sec:simulations}, we take these time series as inputs to our algorithm. Figure~\ref{diag:spitzer3} shows the DWT, using db4 wavelets, of the time series at  7.6971$\mu$m (black lines) with the injected calibration signal over plotted in red. Note that no binning in wavelength was performed before, which marks a significant difference to the HST/NICMOS analysis in \citet{waldmann13}, where a relatively coarse binning was necessary to reduce the Gaussian noise. 

\begin{figure}
\centering
\includegraphics[width=7cm]{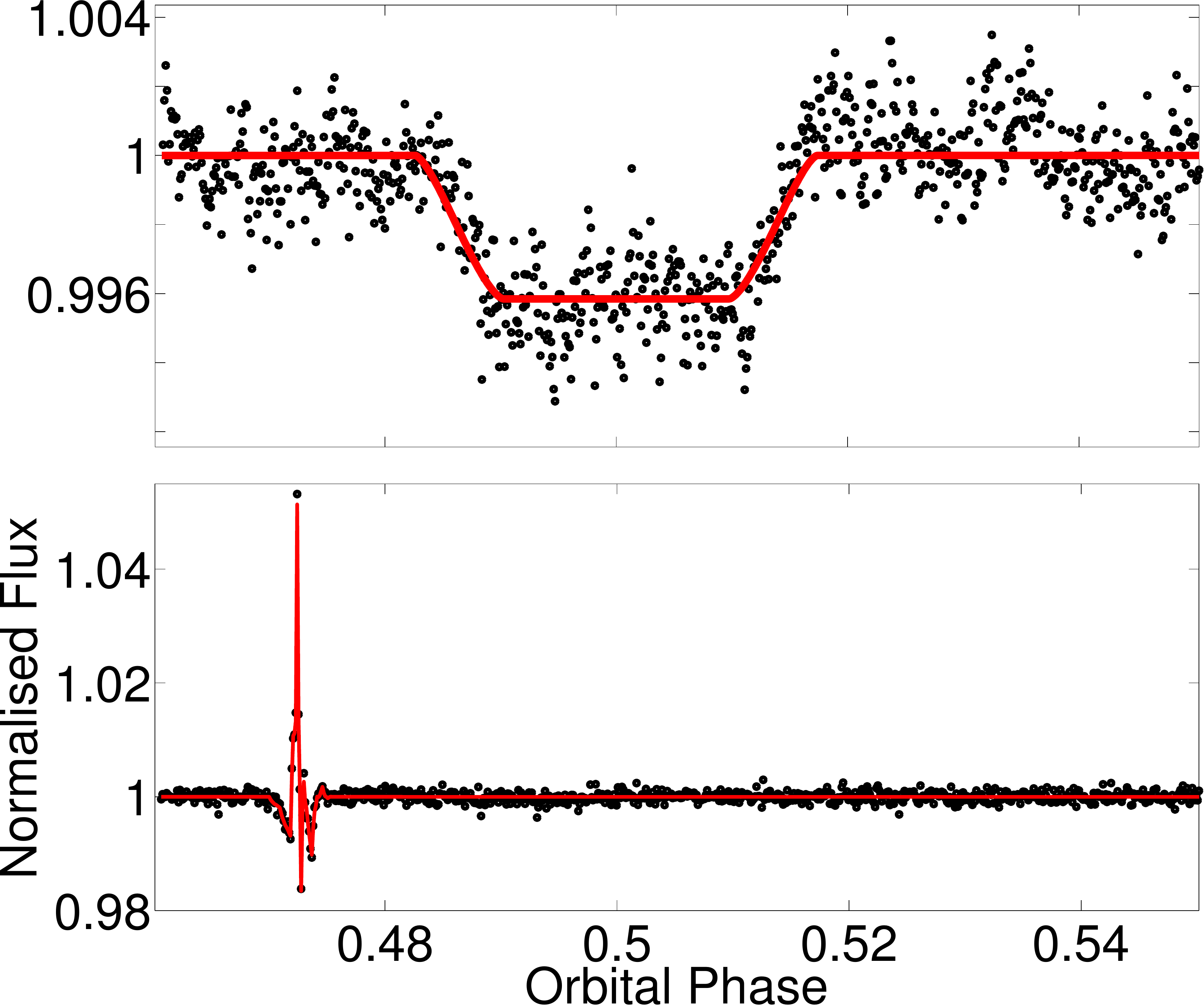}
\caption{TOP: secondary eclipse component at 7.6971$\mu$m (first independent component in figure~\ref{diag:spitzer1}) with \citet{mandel02} model over plotted (red). BOTTOM:  retrieved calibration signal (black dots) with original input calibration signal over plotted (red line). \label{diag:spitzer4}}
\end{figure}

\begin{figure}
\centering
\includegraphics[width=7cm]{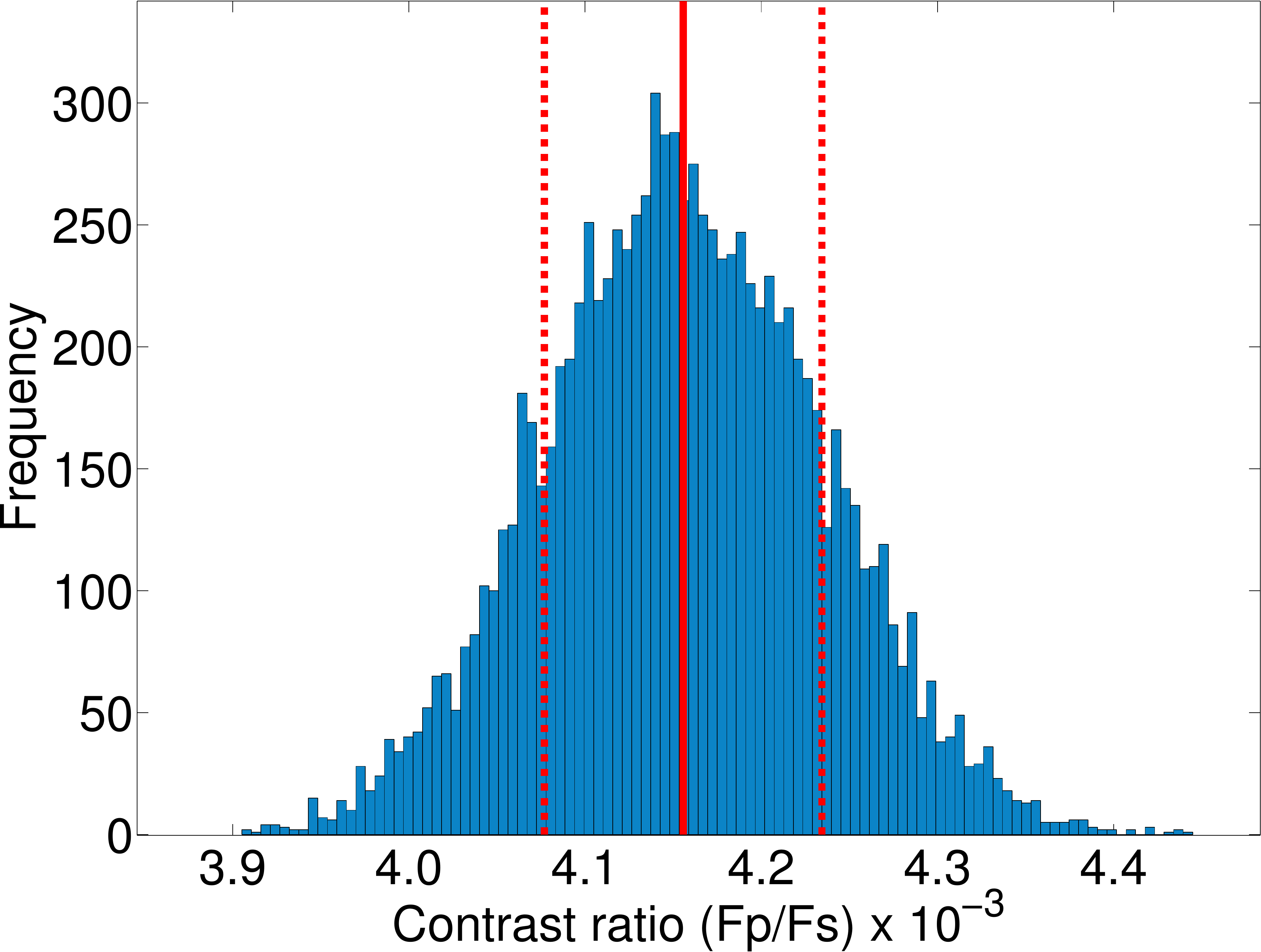}
\caption{MCMC generated posterior distribution of transit depth parameter for lightcurve fit in figure~\ref{diag:spitzer4}. Distribution mean and one sigma bounds are indicated with red continuous and discontinuous lines respectively. \label{diag:spitzer5}}
\end{figure}

We now follow through each individual step as described in the previous sections and obtain the scaled independent components of the data set. Figures~\ref{diag:spitzer1}~\&~\ref{diag:spitzer2} show two observed time series (blue crosses) with the correctly scaled individual independent components (black dots) underneath. The first and second ICs comprise the secondary eclipse signal and the calibration signal respectively. These are also shown in figure~\ref{diag:spitzer4}.  The remaining ICs are instrument or stellar systematic noise. The amplitudes of these systematic components can be seen to be lower toward the blue end of the spectrum (figure~\ref{diag:spitzer1}), which is also evident in a cleaner observed time series, and more pronounced toward the red end of the spectrum (figure~\ref{diag:spitzer2}). In figure~\ref{diag:spitzer4} we over plotted the best fit \citet{mandel02} model using a MCMC fitting routine \citep{waldmann13} with the transit depth as only free parameter. The transit depth posterior distribution is shown in figure~\ref{diag:spitzer5}. The total error is the quadrature sum of $\sigma^{c}_{l=1}$ and the MCMC derived standard error $\sigma_{mc}$. We obtain at 7.6971$\mu$m a planet/star contrast ratio of $Fp/Fs = 0.00415 \pm 0.00015$ which we find consistent with the measurement by \citet{grillmair07}. Figure~\ref{diag:spitzer4} also shows the retrieved calibration signal (black dots) with the original calibration signal over plotted in red. The excellent match between both indicates an adequate signal separation and the correct scaling of the independent components.

\begin{figure}
\centering
\includegraphics[width=7cm]{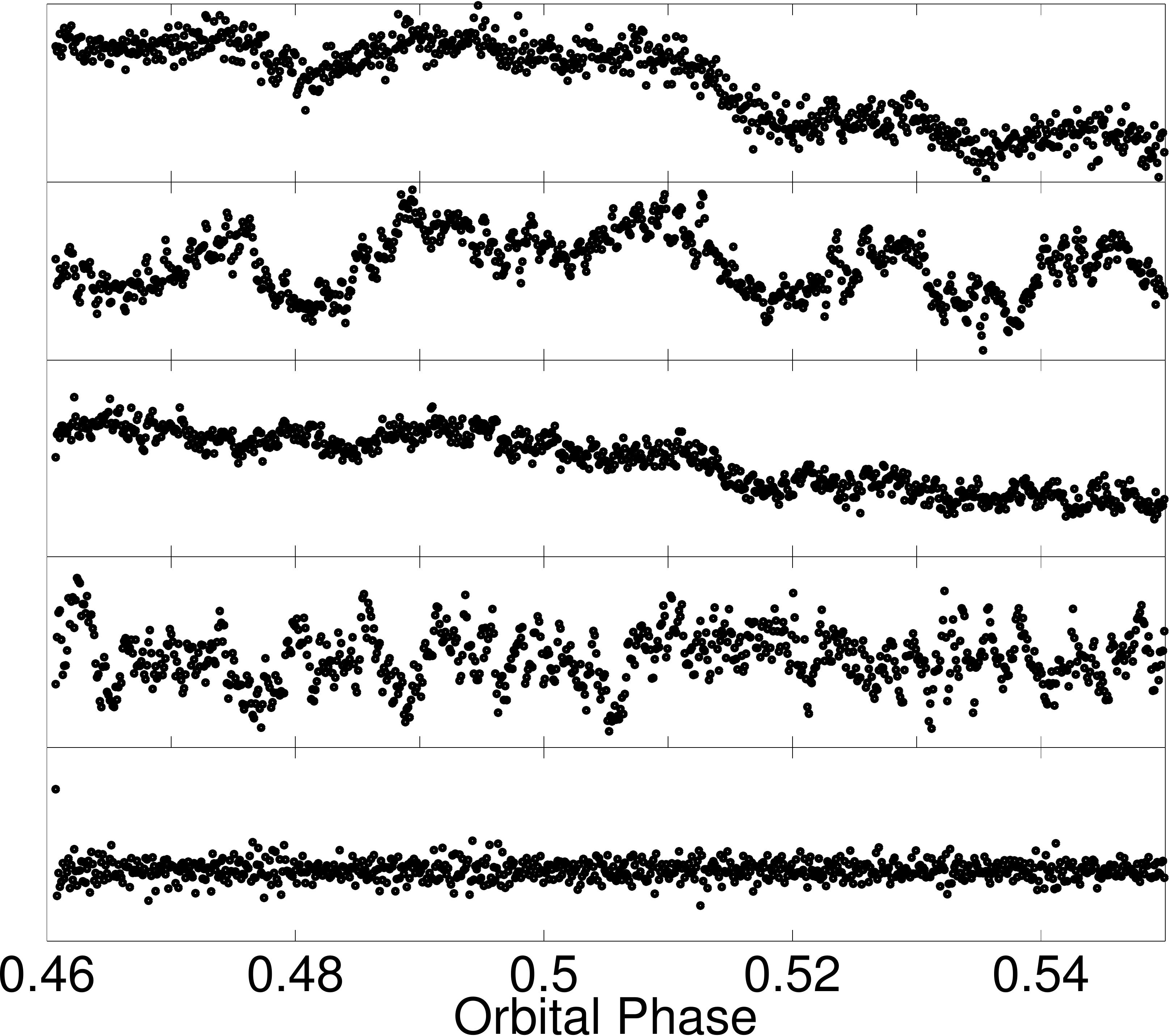}
\caption{First five (unscaled) independent components of ICA analysis performed in the time domain. Given the low signal to noise of the data the convergence of the ICA algorithm is limited and it cannot separate individual sources when directly applied in the time domain. Compare this to figure~\ref{diag:spitzer1} where ICA was performed on the wavelet coefficients and the source signals were separated correctly.  \label{diag:spitzer6}}
\end{figure}

In order to demonstrate the increased efficiency of the proposed \texttt{ACICA} algorithm in the presence of noise, we also show the ICs derived by performing the more `traditional' ICA in the time domain only (figure~\ref{diag:spitzer6}). The components in figure~\ref{diag:spitzer6} are poorly separated and the standard ICA analysis did not converged with traces of the secondary eclipse feature present in three separate components. 

\section{Discussion}

Using \texttt{ACICA}, we can directly retrieve the signal amplitudes of the source components comprising our data. It not only allows us to non-parametrically de-trend low S/N data sets but also allows for a unique insight into the structural make up of the observations. A study of the systematic components and their amplitudes offers a powerful diagnostic on systematic noise behaviour across a detector. As shown in the Spitzer/IRS example, amplitudes of systematic components in the data can vary significantly across the dispersion of a grism. A study of these systematic components may allow, amongst others, for an optimisation of residual flat-fielding errors across a field or the characterisation of wavelength dependent slit-losses of the instrument. As data diagnostic we can hence define the S/N of a time series $k$ in terms of its systematic noise  

\begin{equation}
SNR^{sys}_{k} = \frac{\Delta F_{k}}{\sqrt{\sum_{l,k=1}^{N_{sn}} \sigma^{2}(s^{c}_{sn,l,k})}}
\end{equation}

\noindent where $\sigma^{2}(s^{c}_{sn,l,k})$ is the variance of a given systematic noise component at for the $k^{th}$ time series and $\Delta F_{k}$ is the respective measured transit depth. 

The choice of wavelet family can be an important factor in the convergence properties of the code. For noisy data we find a db4 wavelet base to be the best compromise between slowly varying continues functions and step-like, discrete offsets in the data. Should the data's systematics be dominated by offset patterns, such as those often induced by nodding \citep[e.g.][]{grillmair07}, we find a {\it Haar} wavelet to give a sparser representation of the data than the {\it Daubechies} family. Similarly, high order {\it Daubechies} wavelets or symmetric wavelets ({\it Symlets}) are best suited if one's data is defined by continuous evolving trends.

As mentioned in section~\ref{sec:waveletica}, Gaussian noise can be suppressed in wavelet space by the selective thresholding of detailed coefficients. Whilst true in theory, we found it difficult to implement in practice and great care needs to be taken in the choice of thresholding rule and amplitude to avoid `over cleaning' and distorting of the underlying data. Fortunately we find that convergence properties of the ICA de-convolution are much enhanced by the transformation of the data into wavelet space alone and further cleaning was (in the scope of this analysis) unnecessary.

\section{Conclusion}

In this paper we present a novel approach to the amplitude calibration of independent component analysis (the \texttt{ACICA} algorithm) via the introduction of a sparse calibration signal in orthogonal wavelet space. By transforming observed time series into series of wavelet coefficients, we are able to overcome two main limitations of blind source separation algorithms such as ICA: 1) the convergence of ICA algorithms is strongly impaired by the presence of Gaussian noise in the data. By transforming data to a sparser and more redundant basis we can significantly improve the performance of the ICA de-convolution without otherwise altering the data. 2) In the wavelet domain it is possible to inject an artificial calibration signal of known amplitude into the data without significant or any impact to the original data. With the use of such a calibration signal we can directly determine the individual absolute amplitudes of the derived independent components. This marks a significant improvement over methods such as they are discussed in \citet{waldmann13} where component amplitudes were derived through a regression analysis to existing out of transit baseline data. 

\texttt{ACICA} allows us to de-trend otherwise inaccessible data sets non-parametrically. We demonstrated this using simulations and archival Spitzer/IRS data. It furthermore offers us an unprecedented and unique insight into the morphology of a data set by allowing us to directly map out temporal/wavelength dependent variations of instrumental or stellar noise in the data set. 

Together, these attributes make the algorithm proposed here a highly versatile and powerful tool for exoplanetary time series analysis.

\appendix

\section{The quadrature mirror filter} 
\label{sec:QMF}

A very fast and simple implementation of the DWT for multi-resolution decomposition is by constructing the quadrature mirror transformation matrix. For a Daubechies-4 wavelet we obtain four coefficients comprising $h(t)$ and similarly $g(t)$. Rather than convolving the time series, $x(t)$ with both filters separately and then down-sample, we can also construct a matrix where each odd row contains $h(t)$ and each even row contains $g(t)$ coefficients. This automatically down-samples the data to the new resolution $\varphi+1$. Such a matrix is called a `quadrature mirror filter' (QMF) and equation~\ref{qmf} is an example of such \citep{press07} \\

\begin{equation}
\left [
\begin{array}{ccccccccccc}
c_{0} & c_{1} & c_{2} & c_{3}	&	&	&	& 	&	&	&  \\
c_{3} & -c_{2} & c_{1} & -c_{0}	&	& 	&	& 	&	&	&\\
	&	& c_{0} & c_{1} & c_{2} & c_{3}	&	&	&	& 	& \\
	&	&c_{3} & -c_{2} & c_{1} & -c_{0}&	& 	&	& 	&\\
\vdots & \vdots &	 & 	&	&	& \ddots&	 	&	&	&\\
	&	& 	&	&	&	&	&c_{0} & c_{1} & c_{2} & c_{3}  \\
	&	& 	&	&	&	&	&c_{3} & -c_{2} & c_{1} & -c_{0} \\
c_{2}&c_{3}& 	&	&	&	&	&	 &	 & c_{0} & c_{1}  \\
c_{1}&-c_{0}& 	&	&	&	&	&	 &	 & c_{3} & -c_{2}  \\
\end{array}
\right ]
\label{qmf}
\end{equation}\\

\noindent where the empty spaces denote zeros. To obtain a DWT using this QMF, we multiply the QMF with the column vector containing the time-series on the right.  Note the circular behaviour of the matrix at the bottom, where the wavelet coefficients wrap around to the beginning. This has important consequences as it indicates that the DWT wraps around the data and the transform at the end of the time-series is sensitive to data at the beginning of the time series. This effect can be avoided by adding sufficient zero-valued points to the time series at its beginning and end. This process is also known as `zero-padding'.

%
%
%
%
%
%
%
%
%


\begin{thebibliography}{}


%
%
%
%
%
%
%

%
%

%

%
%
%
%
%

%
%
%
%
%

%
\bibitem[Carter \& Winn (2009)]{carter09} Carter J.A., Winn J.N., 2009, ApJ, 704, 51
%
%
%
%
%
%
\bibitem[Claret (2000)]{claret00} Claret, A., 2000, A$\&$A, 363, 1081
%
%
\bibitem[Comon \& Jutten (2010)]{icabook2} Comon, P., \& Jutten, C., 2010, `Handbook of Blind Source Separation: Independent Component Analysis and Applications', Academic Press
%
\bibitem[Crouzet et al. (2012)]{crouzet12} Crouzet, N., McCullough, P.~R., Burke, C.~J., Long, D., 2012, axXiv: 1210.5275v1
 

\bibitem[Daubechies (1988)]{daubechies88} Daubechies, I., 1988, Commun. Pure Appl. Math, 41, 909

\bibitem[Daubechies (1992)]{daubechies92} Daubechies, I., 1992, `Ten Lectures on Wavelets', Soc. for Industrial and Applied Mathematics
%
%
%
%
%



\bibitem[Donoho (1995)]{donoho95} Donoho,D.L., 1995, IEEE Trans. on. Inf. Theory, 41,3,613



%

%

%
\bibitem[Gibson et al. (2012)]{gibson12} Gibson, N. P., Aigrain, S., Roberts,  S., et al., 2012, MNRAS, 419, 2683
%
%
%
\bibitem[Grillmair et al. (2007)]{grillmair07} Grillmair, C.~J., Charbonneau, D., Burrows, A., et al., 2007, ApJL, 658, L115

\bibitem[Grillmair et al. (2008)]{grillmair08} Grillmair, C.~J., Burrows, A., Charbonneau, D., Armus, L., Stauffer, J., Meadows, V., van Cleve, J., von Braun, K. \& Levine, D., 2008, Nature, 456, 767
%
\bibitem[Haario et al. (2006)]{haario06} Haario, H., Laine, L., Mira, A., Saksman, E., 2006, Statistics and Computing, 16, 339
%
\bibitem[Haario et al. (2001)]{haario01} Haario, H., Saksman, E., Tamminen, J., 2001, Bernoulli, 7, 223
%
\bibitem[Hastings (1970)]{hastings70} Hastings, W.~K., 1970, Biometrika, 57, 97

%
%
%

\bibitem[Houck et al. (2004)]{houck04} Houck, J.~R., Roellig, T.~L., van Cleve, J., et al., 2004, ApJS, 154,18
\bibitem[Hyv\"arinen (1999)]{hyvarinen99} Hyv\"arinen A.,1999, IEEE Trans. on Neural Networks, 10, 626
\bibitem[Hyv\"arinen \& Oja (2000)]{hyvarinen00} Hyv\"arinen A., Oja, E., 2000, Neural Networks, 13, 411
\bibitem[Hyv\"arinen \& Oja. (2001)]{icabook} Hyv\"arinen A., Karhunen J., Oja E., 2001, `Independent Component Analysis',  John Wiley \& Sons Inc. 
\bibitem[Inuso et al. (2007)]{inuso07} Inuso, G., La Foresta, F., Mammone, N., Morabito, F., C., 2007, Int. J. Conf. on Neural Networks, 1524


%
%
%

	
%
%
\bibitem[Koldovsk\'y et al. (2006)]{koldovsky06} Koldovsk\'y Z., Tichavsk\'y P., Oja E., 2006, IEEE Trans. on Neural Networks, 17,1265

\bibitem[Koldovsk\'y et al. (2005)]{koldovsky05} Koldovsk\'y Z., Tichavsk\'y P., Oja E., 2005, Proc.of IEEE/SP 13th Workshop on Stat. Signal Processing
%
%
%
\bibitem[Lin \& Zhang (2005)]{lin05} Lin, J., Zhang, A., 2005, NDT\&E International, 38, 421



%
%
\bibitem[Mammone et al. (2012)]{mammone12} Mammone, N., La Foresta, F., Morabito, F.~C., 2012, IEEE Sensors Journal, 12, 533

\bibitem[Mandel \& Agol (2002)]{mandel02} Mandel, K.and Agol, E., 2002, ApJL, 580, L171


%
%
%
%

%
%
\bibitem[Oja (1992)]{oja92} Oja, E., 1992, Neural Networks, 5, 927

%


\bibitem[Percival \& Walden (2000)]{percival00}Percival,D.B., Walden, A.T.,2000, `Wavelet Methods for Time Series Analysis', Cambridge Uni. Press

%
%
%
\bibitem[Press et al. (2007)]{press07} Press, W.~H., Teukolsky, S.~A., Vetterling, W.~T., Flannery, B.~P., 2007, `Numerical Recipes', Cambridge Uni. Press



%
\bibitem[Riley et al. (2002)]{riley02} Riley, K.~F., Hobson, M.,~P., Bence, S.~J., 2002, `Mathematical Methods for Physics and Engineering', Cambridge Uni. Press 
%
\bibitem[Shannon (1948)]{shannon48} Shannon, C., 1948, Bell System Tech. J., 27, 379
%
%


\bibitem[Stein (1981)]{stein81} Stein, C.M., 1981, Ann. Statist., 9, 6, 1135

%
%
\bibitem[Stone (2004)]{icabook3} Stone, J., V., 2004, `Independent Component Analysis: A tutorial Introduction', A Bradford Book
%
\bibitem[Swain et al. (2008)]{swain08} Swain, M.~R.,Vasisht, G. and Tinetti, G., 2008, Nature, 452, 329
%
\bibitem[Swain et al. (2009a)]{swain09a} {Swain}, M.~R., {Vasisht}, G., {Tinetti}, G., {Bouwman}, J., 
	{Chen}, P., et al , 2009, ApJL, 690, L114
\bibitem[Swain et al. (2009b)]{swain09b} {Swain}, M.~R., {Tinetti}, G., {Vasisht}, G., {Deroo}, P., 
	{Griffith}, C., et al., D., 2009, ApJ, 704, 1616
%
\bibitem[Swain et al. (2011)]{swain11} {Swain}, M.~R. and {Deroo}, P. and {Vasisht}, G., 2011, IAU Symposium, 276, 148


\bibitem[Tichavsk\'y et al. (2006a)]{tichavsky06} Tichavsk\'y P., Doron E., Yeredor A., Gomez-Herrero G., 2006, Proc. EUSIPCO-2006
\bibitem[Tichavsk\'y et al. (2006b)]{tichavsky06b} Tichavsk\'y P., Doron E., Yeredor A., Nielsen J., 2006, Proc. EUSIPCO-2006
%
\bibitem[Tichavsk\'y et al. (2008)]{tichavsky08}Tichavsky, P., Koldovsky, Z., Yeredor, A., Gomez-Herrero, G. , Doron, E., 2008, IEEE Transactions on Neural Networks, 19, 421

\bibitem[Tessenyi et al. (2012)]{tessenyi12} Tessenyi, M., Ollivier, M., Tinetti, G., et al., 2012, ApJ, 746, 45


%
%
%
  

\bibitem[Thatte et al. (2010)]{thatte10} {Thatte}, A. and {Deroo}, P. and {Swain}, M.~R., 2010, A\&A, 523, 35

%
%
%
%
	
%



\bibitem[Waldmann (2012)]{waldmann12b} Waldmann, I.~P., 2012, ApJ, 747, 12

\bibitem[Waldmann et al. (2013)]{waldmann13} Waldmann, I~P., Tinetti, G., Deroo, P., et al., 2013, ApJ accepted
%
%
%
%
\bibitem[Yeredor (2000)]{yeredor00} Yeredor A., 2000, IEEE Sig. Proc. Letters, 7, 197
%


\end{thebibliography}
\end{document}